\documentclass[a4paper,twocolumn,11pt]{quantumarticle}
\pdfoutput=1
\usepackage[utf8]{inputenc}
\usepackage[english]{babel}
\usepackage[T1]{fontenc}
\usepackage{amsmath}
\usepackage{hyperref}

\usepackage{tikz}
\usepackage{lipsum}
\usepackage[numbers]{natbib}
\usepackage{amssymb}
\newcommand{\ie}{\textit{i.e.,\ }}

\begin{document}

\title{Certification of genuine non-Gaussian entanglement}

\author{Lukáš Lachman}
\affiliation{Department of Optics, Palack\' y University, 17. listopadu 12, 77146 Olomouc, Czech Republic}
\email{lachman@optics.upol.cz}

\author{Carlos E. Lopetegui-Gonz\'alez}
\affiliation{Laboratoire Kastler Brossel, Sorbonne Universit\'e, CNRS, ENS-Universit\'e PSL, Coll\`ege de France, 4 Place Jussieu, 75005 Paris, France}

\author{Massimo Frigerio}
\affiliation{Laboratoire Kastler Brossel, Sorbonne Universit\'e, CNRS, ENS-Universit\'e PSL, Coll\`ege de France, 4 Place Jussieu, 75005 Paris, France}

\author{Mattia Walschaers}
\email{mattia.walschaers@lkb.upmc.fr}
\affiliation{Laboratoire Kastler Brossel, Sorbonne Universit\'e, CNRS, ENS-Universit\'e PSL, Coll\`ege de France, 4 Place Jussieu, 75005 Paris, France}

\maketitle

\begin{abstract}
 Entanglement is a key resource for many quantum applications. Understanding fundamental properties of entangled states is an important step towards their practical exploitation. We characterize entanglement in the context of Gaussian and non-Gaussian processes  and identify entangled states that cannot be produced by any Gaussian evolution acting on separable states. This distinction exposes intrinsic limitations of Gaussian operations and highlights the role of non-Gaussian operations as an important resource. We apply this framework to develop a certification method that connects entanglement theory with quantum non-Gaussianity - an advanced quantum feature that is essential for several quantum applications. Our approach tailors the certification to experimentally relevant states that can be produced in current quantum optics experiments. We demonstrate this certification for recognizing the genuine non-Gaussian entanglement of various entangled Fock states and hybrid entangled states.
\end{abstract}

\section{Introduction}
The rapid development of quantum technologies promises significant progress in solving practical tasks that rely on quantum resources \cite{Gisin2007,Acin2018}. Considerable technological effort is put into controlling and engineering various bosonic states and their use in applications such as quantum sensing \cite{Wolf2019,McCormick2019,Wang2019,Pan2025,Rahman2025}, quantum communication \cite{Pirandola2020,Kumar2025}, quantum simulations \cite{Daley2022} or quantum computations \cite{Menicucci2014,Bravyi2022,Madsen2022,AbuGhanem2025,Monbroussou2025,Xanadu_2025,konno_logical_2024,Kudra2022,fluhmann_encoding_2019}. Achieving these goals, however, requires a detailed understanding of the physical resources needed to produce suitable bosonic states. One of the necessary resources for realizing these tasks beyond what would be possible with classical systems is quantum non-Gaussianity \cite{mari_2012,Chabaud2023,Walschaers2021}. Many technological applications require us to go beyond Gaussian resources, making certification of the realization of such non-Gaussian features a crucial task. A theoretical framework for the study of quantum non-Gaussianity has been established \cite{Walschaers2021,Lachman2022} and further refined through hierarchical classifications of non-Gaussian states \cite{Lachman2019,Chabaud2020}. 

As a prototypical example, Fock states and finite superpositions of them have long been recognized as key non-Gaussian resources for bosonic quantum technologies, including sensing, simulation, and fault-tolerant information processing. Their role is now understood more precisely within the quantum non-Gaussian hierarchy \cite{Podhora2022,Chabaud2023}. Beyond providing a basic distinction between Gaussian and non-Gaussian states, these hierarchies relate non-Gaussianity to the minimal discrete resources, such as photon additions or Fock-state components, required to generate a state using Gaussian operations. One such classification is provided by the \emph{stellar rank}, quantifying the minimal number of photons in the core state, i.e. the finite Fock superposition from which the target state can be obtained by a Gaussian unitary \cite{Lachman2019,Chabaud2020,Walschaers2021}. In this sense, the hierarchy does not just detect non-Gaussianity, but resolves different levels of non-Gaussian resourcefulness.

Quantum coherence constitutes another central quantum resource, especially in quantum computing and bosonic error-correction architectures \cite{Streltsov2017,Michael2016}. Recently, the framework of quantum non-Gaussianity has been extended to account for distinctive off-diagonal elements (coherences) in the Fock basis, leading to the notion of quantum non-Gaussian coherence \cite{Asenbeck2025,Lachman2025}. Unlike the stellar hierarchy, which classifies states through the minimal superposition of Fock states required for their generation, these new approaches target specific coherences and certify whether they can arise from Gaussian states or Gaussian processes. They therefore provide a complementary, context-dependent hierarchy for genuinely non-Gaussian coherent features.

A natural next step is to develop an equally precise classification for non-Gaussian entanglement. Entanglement is a genuinely multipartite manifestation of quantum coherence and a central resource for quantum information processing \cite{Piveteau2022}, metrology \cite{pezze_quantum_2018}, and computation \cite{Streltsov2017,Nagata2007,Ou2020,Hong2021,Arute2019}. Recent progress has identified stronger forms of bosonic entanglement that already go beyond standard Gaussian mode mixing. In particular, mode-intrinsic entanglement characterizes correlations that cannot be removed by passive linear-optical transformations, i.e. by mode basis changing implemented with beam-splitters and phase shifters \cite{sperling_mode-independent_2019,chabaud_holomorphic_2022,walschaers_statistical_2017,Geometric_2025}. This notion, however, does not exhaust the structure of entangled Gaussian evolutions: active Gaussian processes such as parametric amplification or two-mode squeezing can generate entangled states that are not naturally captured within a beam-splitter-based picture. To address this broader boundary, the notion of genuine non-Gaussian entanglement has recently been introduced for entangled states that cannot be generated from separable inputs by Gaussian protocols \cite{Zhao2025}. Together, these developments indicate that the classification of non-Gaussian entanglement is only beginning to emerge: mode-intrinsic entanglement isolates correlations robust against passive-linear-optics mode basis changes, while genuine non-Gaussian entanglement identifies correlations that cannot be achieved with any entangling Gaussian operation. Establishing the relation between these notions is therefore an important step toward a unified resource theory of non-Gaussian entanglement and toward identifying which forms of entanglement are truly required in future bosonic quantum technologies.
\begin{figure*}[t]
	%\vspace{-1cm}
	\includegraphics[width=16cm]{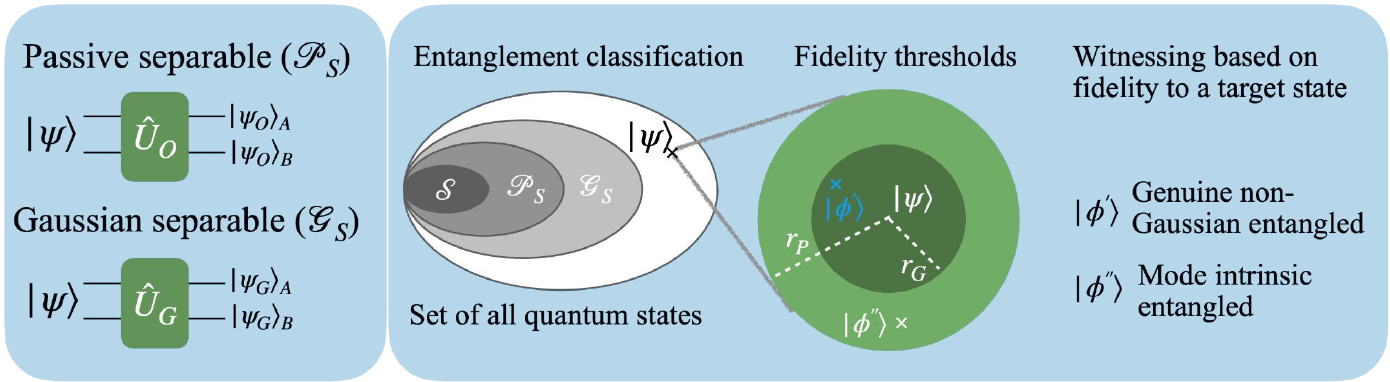}
	%\vspace{-1cm}
	\caption{We tackle the question of identifying whether a given state has entanglement that cannot be undone by Gaussian or linear optics (passive) operations. The set of states that can be made separable under Gaussian unitaries ($\mathcal G_S$) is a convex set, and contains the (also convex) set of passive separable states ($\mathcal P_S$). If a state does not belong to either of these sets, then it is said to be genuinely non-Gaussian entangled. Since these form an open set, there is an open neighborhood around each state (in the topology induced by the fidelity) containing only states that cannot be disentangled using Gaussian operations (ball depicted with radius $r_G$), and a larger one in which only states that are not passive separable can be found (identified with $r_P$). This robustness can be used as a witness of genuine non-Gaussian entanglement and mode-intrinsic entanglement based on the fidelity to the target state. If the fidelity is greater than the passive (Gaussian) separable we can guaranty that the state is mode-intrinsic (genuinely non-Gaussian) entangled.}\label{fig1} 
\end{figure*}
Previous efforts on the practical certification of these strong forms of non-Gaussian entanglement were limited to the certification of mode-intrinsic entanglement through metrological techniques \cite{Lopetegui2025}. Such an approach has the disadvantage that it only witnesses metrologically useful entanglement, which implies that some instances of non-Gaussian entanglement, relevant for quantum information processing, may remain undetected. 

In the present work, we derive a new method to certify quantum non-Gaussian entanglement of states, based on their proximity to well-defined target states. We provide an efficient approach allowing us to derive thresholds needed for the certification based on the fidelity to such target states. We apply this method on entangled states relevant for prospective quantum applications, and for near term quantum optics experiments. Further, we analyze the realistic conditions enabling the certification.

\section{Gaussian and passive separability}
We focus here only on pure 2-mode states. A pure separable state can be written as $|\phi\rangle=|\phi_1\rangle_1 |\phi_2\rangle_2$ where the subscripts $1$ and $2$ represent the mode indexes. A pure entangled state is defined as a state that is not factorized.  We further classify entangled states by the physical resources required for their generation in the framework of Gaussian operations \cite{Bloch1962,Braunstein2005}. 
%Any Gaussian operation admits a Bloch-Messiah decomposition \cite{Bloch1962,Braunstein2005}.
In particular, we call a 2-mode entangled state \emph{Gaussian separable} if the state can be obtained by applying a generic Gaussian operation $U_G$ to a separable state. Given the unbounded and non-energy preserving character of the squeezing operation, it is also common to consider \emph{passive separable states} as a subclass of Gaussian entangled states that can be generated by acting only with a passive linear optics operation on a separable state. As an important remark, this distinction truly arises only for non-Gaussian states, as all Gaussian states are always passively separable \cite{Walschaers2021}. 
Let us notice that only some Gaussian operations can generate entanglement from separable states. We say that $U_G$ is an entanglement-generating Gaussian unitary operation if the state $U_G |\psi_1\rangle_1|\psi_2\rangle_2$ is entangled for at least one separable state $|\psi_1\rangle_1 |\psi_2\rangle_2$ undergoing the Gaussian evolution $U_G$. Interference at a beam-splitter provides a basic example of entanglement-generating passive Gaussian operation, whose unitary operator is $U_{BS}=\exp(\tau a^{\dagger}_1 a_2-\tau^* a_1 a_2^{\dagger})$. Generally, any entanglement-generating Gaussian operation can be parameterized as:
\begin{equation}\label{GaussianEntanglement}
U_G=U_{BS}(\tau_1)S_{1,2}(\xi)U_{BS}(\tau_2),
\end{equation}
where $S_{1,2}(\xi)=\exp\left(\xi a_1^{\dagger}a_2^{\dagger}-\xi^* a_1 a_2\right)$ is a two-mode squeezing unitary operator. In the Appendix, we prove that Eq.~(\ref{GaussianEntanglement}) determines the full set of entanglement-generating Gaussian operations.

\section{Genuine non-Gaussian entanglement}
\label{sec:method}
Entanglement generated by Gaussian operations does not cover all physical situations. We say that a pure 2-mode state $\vert \psi \rangle$ possesses \emph{genuine non-Gaussian entanglement} if $|\psi\rangle \neq U_G |\phi_1\rangle_1 |\phi_2\rangle_2$ for any Gaussian unitary operation $U_{G}$ and any initial pure separable state $|\phi_1\rangle_1 |\phi_2\rangle_2$. Analogously to passive separability, we further call \emph{mode-intrinsic entangled} those two-mode states such that $|\psi\rangle \neq U_{BS} |\phi_1\rangle_1 |\phi_2\rangle_2$. \par 
As a formal extension beyond pure states, a mixed state $\rho$ is genuinely non-Gaussian entangled (resp. mode-intrinsically entangled) if it cannot be expressed as a mixture of Gaussian entangled (resp. passively separable) states. In \cite{Geometric_2025}, a discussion was outlined about the difficulty of properly choosing an extension of the concept of passive or Gaussian separability to mixed states. Ultimately, it depends on the kind of resource theory one would like to construct. Here, we constrain ourselves to a convex resource theory and choose to certify non-Gaussian entanglement that cannot be reproduced from convex mixtures of passive (Gaussian) separable states. Notice that this does not require that there is a specific and unique mode basis change (or Gaussian unitary transformation) that disentangles the state, as the different elements in the mixture may be separable in different bases. This is in contrast to the method developed in \cite{Lopetegui2025}, which witnesses mode-intrinsic entanglement by asking the question of whether there is a single beam splitter operation that disentangles the state.  

In a convex resource theory every resourceful state is surrounded by a region in the state space, where only other resourceful states are present. This implies that there is some threshold for the fidelity of a resource-free state to this \textit{target} state \cite{brandao_separable_2004}. Therefore, to certify genuine non-Gaussian entanglement of a given state, we can determine its fidelity to a  chosen target state $|\psi_t\rangle$ and compare it with a threshold $\mathcal{T}_{G,|\psi_t\rangle}$ given by the maximal fidelity between a state produced by Gaussian operations acting on a separable state, \emph{i.e.}
\begin{equation}\label{eq:threshold_definition}
    \mathcal{T}_{G,|\psi_t\rangle}=\max_{|\phi_1\rangle_1,|\phi_2\rangle_2, U_G} |\langle \psi_t|U_G |\phi_1\rangle_1 |\phi_2\rangle_2|^2.
\end{equation}
The certification is successful if the fidelity obtained for the measured state exceeds the threshold $\mathcal{T}_{G,|\psi_t\rangle}$.
Analogously, certification of the mode-intrinsic entanglement compares the fidelity of the measured state with the threshold $\mathcal{T}_{O,|\psi_t\rangle}$ corresponding to the maximal fidelity between the target state $|\psi_t\rangle$ and any passive separable state.
  \par 
The task is to perform the maximization leading to a corresponding threshold. To carry that out, we first maximize the fidelity over separable states $|\phi_1\rangle_1|\phi_2\rangle_2$ with a fixed Gaussian operation.We restrict the states $|\phi_i\rangle_i$ with $i=1,2$ and consider them as superpositions of Fock states up to Fock state $|n\rangle$, i.e. $|\phi_i\rangle_i=\sum_{k=0}^n c_{i,k} |k\rangle_i$, and introduce the maximum
\begin{equation}
	\widetilde{\mathcal{T}}_{|\psi_t\rangle,n}(U)=\max_{|\phi_1\rangle_1,|\phi_2\rangle_2} |\langle \psi_t|U|\phi_1\rangle |\phi_2\rangle|^2,
\end{equation}
where $U$ is a fixed Gaussian operation and $n$ corresponds to the maximal Fock state in the superposition of the states $|\phi_i\rangle_i$. Notice that when $U$ is a passive Gaussian operation the maximum value of $n$ is dictated by the target state, provided that it is a \emph{core state}. When the more general case of Gaussian separability is considered and squeezing is allowed, instead, one has to further maximize the final result by taking larger and larger values of the cutoff $n$ until numerical convergence is achieved.

Let us now introduce a matrix $M$ such that, for any two Fock states $\vert k\rangle_{1}, \vert l \rangle_{2}$ ($k,l \leq n$) we have: 
\begin{equation}
	\begin{aligned}
		M_{kl} \ = \ \langle \psi_t|U| k\rangle_1|l\rangle_2
	\end{aligned}
\end{equation}
If we then call $\boldsymbol{u}$ the vector with entries $u_{k} = \langle \phi_{1} \vert k \rangle_{1}$ and, analogously, $\boldsymbol{v}$ the vector with entries $v_k = \langle k \vert \phi_{2} \rangle_{2}$, we have that:
\begin{equation}
   \widetilde{\mathcal{T}}_{|\psi_t\rangle,n}(U) \ = \ \max_{\boldsymbol{u},\boldsymbol{v} \in \mathcal{S}^n} \vert \boldsymbol{u}^\dagger M \boldsymbol{v} \vert^2
\end{equation}
where $\mathcal{S}^{n}$ is the unit sphere in $\mathbb{C}^n$. This expression is well-known to be equal to the square of the spectral norm of the matrix $M$:
\begin{equation}
  \max_{\boldsymbol{u},\boldsymbol{v} \in \mathcal{S}^n} \vert \boldsymbol{u}^\dagger M \boldsymbol{v} \vert^2 \ = \ \max \mathrm{Spec}(M^\dagger M)
\end{equation}
To obtain the threshold $\mathcal{T}_{G,|\psi_t\rangle}$ or $\mathcal{T}_{O,|\psi_t\rangle}$, we still need to maximize $\widetilde{\mathcal{T}}_{|\psi_t\rangle,n}$ over the Gaussian evolution $U_G$ or the unitary operator $U_{BS}$, respectively. We can recast the optimization in \eqref{eq:threshold_definition} as 
\begin{equation}
    \mathcal{T}_{G(O),|\psi_t\rangle}=\max_{U_{G(O)}}\left[ \text{maxSpec}(M^\dagger M)\right].
\end{equation}
 For the case of two modes, both passive and general Gaussian unitaries can be characterized with a small number of parameters. In the case of passive unitaries one single mode phase shift and one real beam splitter are enough, which are parametrized by angles $\{\phi_1,\theta\}$. In the case of gaussian unitaries, following equation \eqref{GaussianEntanglement}, three complex parameters are sufficient to cover all relevant entangling unitaries. The optimization to be performed is non-convex, which makes it relatively complicated as local minima have to be avoided. Moreover, it is hard to compute gradients over the relevant parameters. This means that the optimization has to be performed through some form of non-gradient based exploration of the landscape. One such form of exploration is provided by a genetic algorithm like the Covariance Matrix Adaptation evolution strategies (CMA-ES) \cite{Hansen2006}, which, provided a carefull choice of hyperparameters for the optimization, lead to efficient and robust implementations of optimizations on continuous domains. We emphasize that in the case of the passive separable example there is no need for an actual optimization and a full scan of the parameter space can be done sufficiently fast, and be more instructive. In the case of Gaussian separability, implementing the genetic algorithm is required. \par

\begin{figure}[t]
	%\vspace{-1cm}
	\includegraphics[width=8cm]{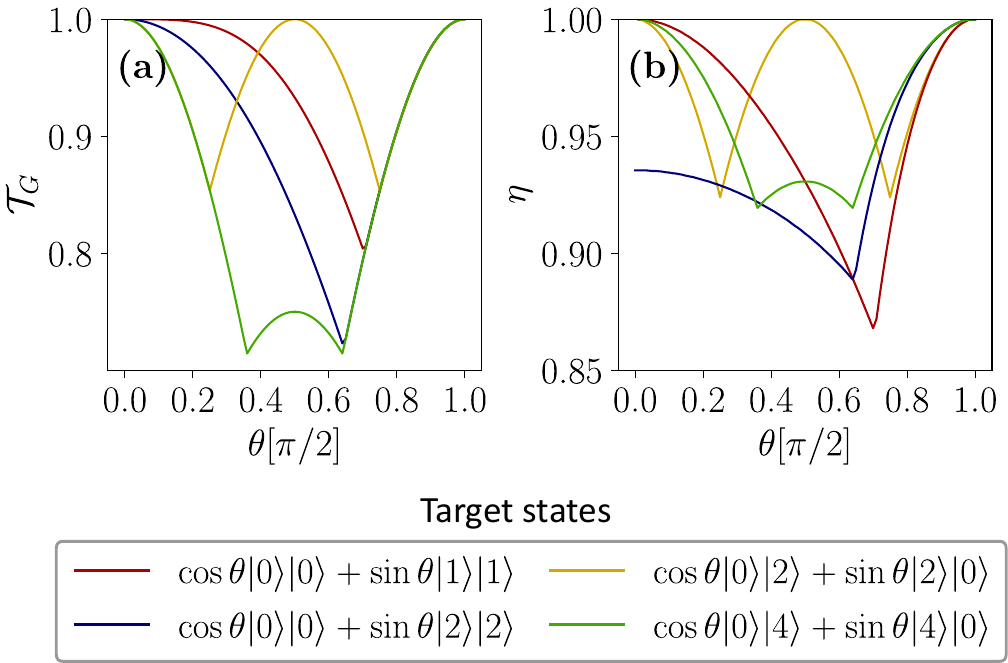}
	%\vspace{-1cm}
	\caption{a) Thresholds for the genuine non-Gaussian entanglement relying on the fidelity with target states $\cos \theta |0\rangle|0\rangle+\sin \theta|1\rangle|1\rangle$ (red line), $\cos \theta |0\rangle|0\rangle+\sin \theta|2\rangle|2\rangle$ (blue line), $\cos \theta |0\rangle|2\rangle+\sin \theta|2\rangle|0\rangle$ (yellow line) and $\cos \theta |0\rangle|4\rangle+\sin \theta|4\rangle|0\rangle$ (green line). Fidelity of a state with a target state surpassing the corresponding threshold certifies the genuine non-Gaussian entanglement. b) Sensitivity of the characterization of genuine non-Gaussian entanglement to losses characterized by minimal transmission efficiency $\eta$ of a lossy channel for which the fidelity to a state entering the channel is above the threshold for the certification of genuine non-Gaussian entanglement using this state as a target. A line of a given color shows the minimal $\eta$ for a target state giving a threshold with the corresponding color in a).}\label{fig2}
\end{figure}

\section{Results}
\label{sec:results}
In this section, we apply the procedure described so far to several target states of interest. The power of this approach lies in using only the fidelity with a target state for certification of mode-intrinsic or genuine non-Gaussian entanglement, circumventing the requirement for full quantum state reconstruction. This makes this certification method versatile and adaptable to an arbitrary targeted experimental outcome. 
\par
To demonstrate this certification, we consider first target states that correspond to finite superposition of Fock states, that may be of interest for platforms that deal with single photons for example. Then we will consider \textit{hybrid} states, that combine DV (single photon level) encoding, and CV encoding, for example including superpositions of tensor products of single photons and coherent states. Finally, we will consider purely continuous variables examples, particularly through the case of single and two photon subtracted states, of immediate interest for experimental implementations. \par

  %\begin{figure*}[t]
	%\vspace{-1cm}
	%\includegraphics[width=8cm]{fidelity_0011_vs_losses.pdf}\includegraphics[width=8cm]{fidelity_0022_vs_losses.pdf}
	%\vspace{-1cm}
	%\caption{Behavior of the fidelity with respect to the pure target state, under the effect of losses for (left) $|\psi_{1,1}\rangle$ and (right) $|\psi_{1,2}\rangle$, for $\theta=0.68$ and $0.65$ respectively. In gray  the region for which genuine non-Gaussian entanglement can be certified through the fidelity to the pure target state. In both cases this is possible for losses up to more than $10\%$.}
    %\label{fig:fidelities_00nn}
%\end{figure*}

\subsection{Finite superpositions of Fock states}
  Entanglement of finite superpositions of multimode Fock states is a resource for different quantum information processing tasks \cite{PsiQuantum,Quandela2023,hoven2025,Namkung_2024}. In quantum optics, these states are usually prepared conditionally, by merging single-photon states and postselecting on photon counting \cite{aralov2025,Chen_2024}.    \par 
  The entangled state $(|0\rangle|1\rangle+|1\rangle|0\rangle)/\sqrt{2}$ is the simplest example and it results from the interference between the Fock state $|1\rangle$ and the vacuum state $|0\rangle$ at a beam-splitter, therefore it is passive separable. In contrast, the state $\cos \theta |0\rangle|0\rangle+\sin \theta |1\rangle|1\rangle$ is neither passive separable nor Gaussian separable for any $\theta \in (0,\pi/2)$. More generally, any state $|\psi_{1,n}(\theta)\rangle=\cos \theta |0\rangle|0\rangle+\sin \theta |n\rangle|n\rangle$ with $n\geq 1$ is not Gaussian separable. Fig.~\ref{fig2} a) depicts corresponding thresholds derived for fidelity with the state $|\psi_{1,n}(\theta)\rangle$ with $n=1$ or $n=2$ and various $\theta$. 
We consider as well states of the form $|\psi_{2,n}(\theta)\rangle=\cos \theta |0,2n\rangle+\sin\theta |2n, 0\rangle$, for $n=1$ and $n=2$. Although the Hong-Ou-Mandel state $(|2\rangle|0\rangle+|0\rangle|2\rangle)/\sqrt{2}$ is passive separable, the unbalanced superposition $\cos \theta |2\rangle|0\rangle+\sin \theta|0\rangle|2\rangle$ with $\theta\neq \pi/4$ exhibits genuine non-Gaussian entanglement. The corresponding threshold is depicted in Fig.~\ref{fig2} a), alongside that for the choice of $n=2$, which is the NOON state with $N=4$ \cite{Kok2002}.
\par
We analyse the sensitivity of this certification to losses by providing minimal transmission efficiency of a lossy channel that keeps the fidelity between a target state affected by the lossy channel and the lossless target state above a corresponding threshold. Fig.~\ref{fig2} b) illustrates this sensitivity to losses for the introduced states $|\psi_{1,n}(\theta)\rangle$ and $|\psi_{2,n}(\theta)\rangle$. It demonstrates requirements of this certification imposed on realistic states. The most striking feature observed is the loss tolerance for small values of $\theta$ for the state $\cos \theta |0\rangle|0\rangle+\sin \theta|1\rangle|1\rangle$. Despite the high value of the threshold, the amount of losses required to make the fidelity to the state entering the loss channel smaller than the threshold for certification is relatively high. Yet, this does not immediately imply that a practical certification for these states would be possible since, once error bars are considered, the difference between the computed fidelity and the threshold would not necessarily be statistically significative. 
  %In Fig.~\ref{fig:fidelities_0nn0}, the tolerance of the certification to a symmetric loss channel, for the choice of $\theta$ that provides the best possible threshold, for the states $|\psi_{2,n}(\theta)\rangle=\cos \theta |0,2n\rangle+\sin\theta |2n, 0\rangle$, for $n=1$, and $n=2$, and the choices of $\theta$ that provide the highest threshold, which are $\theta=0.23 \pi/2$ and $\theta=0.37 \pi/2$ respectively. In the first case, around $6 \%$ losses can be tolerated before the certification is no longer possible with the ideal pure state as the target, while for $n=2$ the tolerance is around $8 \%$ losses. Similar behaviors can be observed for NOON states, \ie for $\theta=\pi/4$ and $N\geq 3$.

%\textcolor{red}{As we discussed on Wednesday we can put together in a single Fig the plots in Fig2 a), Fig 2 d) and Fig 2 f), and leave out for the moment the Fig 2 b and c) and the plots currently in Fig. 3. and 4.}
  \par

 \subsection{Hybrid states}
 Here, we study the genuine non-Gaussian entanglement of the hybrid states \cite{Darras2022} $|\mbox{hybrid}_1\rangle=\cos \theta |0\rangle |\mbox{cat}_+\rangle+\sin \theta|1\rangle |\mbox{cat}_-\rangle$ and $|\mbox{hybrid}_2\rangle=\cos \theta |0\rangle |\mbox{cat}_-\rangle+\sin \theta|1\rangle |\mbox{cat}_+\rangle$, where $|\mbox{cat}_{\pm}\rangle\propto|\alpha\rangle\pm |-\alpha\rangle$ represents the odd or even cat state. Such states are of interest for hybrid information processing, at a time where interconnection between different types of encodings for quantum information processing tasks becomes increasingly relevant \cite{Lee2024}.  
 \par
\begin{figure}[t]
	%\vspace{-1cm}
	\includegraphics[width=8cm]{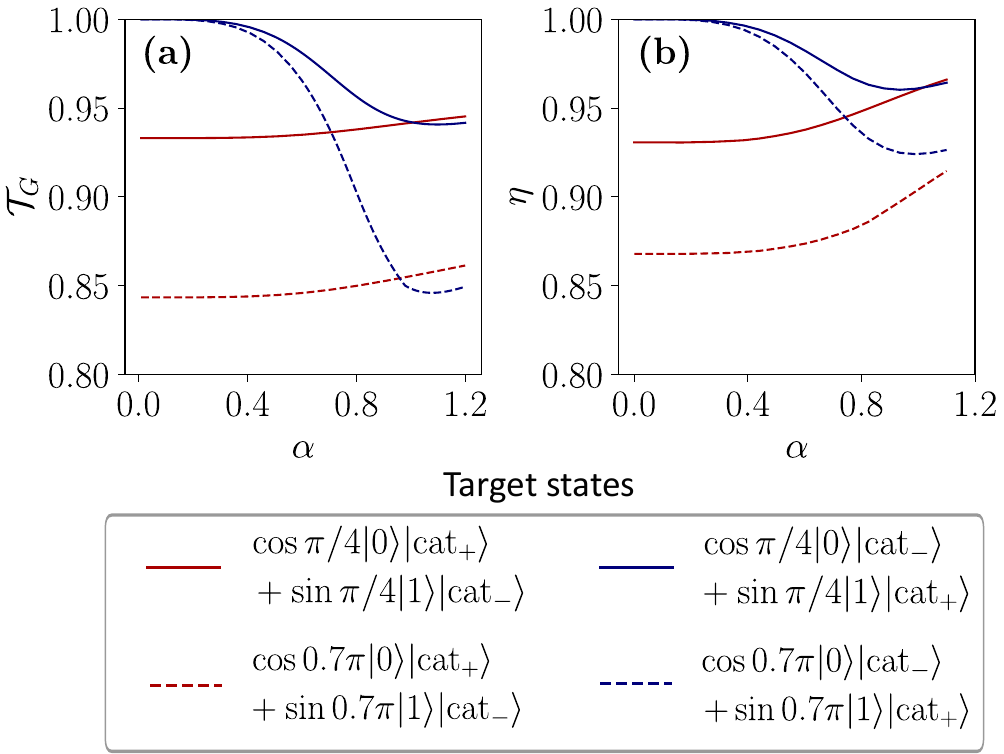}
	%\vspace{-1cm}
	\caption{a) Thresholds for the genuine non-Gaussian entanglement derived for the fidelity with the hybrid states $\cos \theta |0\rangle|\mbox{cat}_+\rangle+\sin \theta|1\rangle|\mbox{cat}_-\rangle$ (red lines), $\cos \theta |0\rangle|\mbox{cat}_-\rangle+\sin \theta|1\rangle|\mbox{cat}_+\rangle$ (blue lines) with $|\mbox{cat}_{\pm}\rangle\propto |\alpha\rangle \pm |-\alpha\rangle$. Whereas the solid lines depict the case $\theta=\pi/4$, the dashed ones presents $\theta=0.7 \pi$ b) Sensitivity of the witness of genuine non-Gaussian entanglement to losses characterized by minimal transmission efficiency $\eta$ of a lossy channel for which the fidelity to the state entering the channel is above the threshold for certification. A line of a given color shows the minimal $\eta$ for the corresponding hybrid state in a).}
    \label{fig3}
\end{figure}

 The thresholds for these hybrid states are shown in Fig.~\ref{fig3} a). In the limit $\alpha \rightarrow 0$ of the state $|\mbox{hybrid}_1\rangle$, we obtain a threshold identical to the state $\sin \theta |0\rangle|0\rangle+\cos \theta |1\rangle|1\rangle$. Whereas the state $|\mbox{hybrid}_2\rangle$ is Gaussian separable only in the limit $\alpha \rightarrow 0$, the case $\alpha>0$ leads to a beatable threshold. We show in Fig.~\ref{fig3} b) the loss tolerance for the certification of genuine non-Gaussian entanglement on those states as a function of their spacing $\alpha$. 
 %the non-Gaussian entanglement of some instances of the states of the kind $|\text{hybrid}_2\rangle$, using as target states their pure state versions. We show two choices of the balancing parameter $\theta=\{\pi/4,0.7\pi/2\}$, and for each of them consider a small value of the spacing $\alpha=0.05$, and a large one $\alpha=1.1$. 
The best tolerance is observed in the unbalanced superposition, \ie for the case in which $\theta=0.7 \pi/2$, for low values of the spacing $\alpha$, with a threshold of losses $\sim 13.5\%.$  
%\textcolor{red}{We can put Fig.2 e) here together with the figures for the loss tolerance of the hybrid states.}

\subsection{Photon subtracted states}
In this section, we consider the case of photon subtracted states, of interest for current optical realizations of non-Gaussian states, which thus offer an immediate test-bed for the ideas discussed in this paper. We consider a state $|\psi_m\rangle$ produced by $m$-photon subtraction on a Gaussian state, which leads to \cite{Barral2024}
\begin{equation}\label{modelSt}
\begin{aligned}
     &|\psi_m\rangle =\\
     & \Pi_{k=0}^m \left(a_1\cos \phi_k+a_2 \sin \phi_k\right)S_1(\xi_1)S_2(\xi_2)|0\rangle |0\rangle,
\end{aligned}
\end{equation}
where $S_i(\xi)=\exp\left[\xi\left(a^{\dagger}\right)^2-\xi^* a^2\right]$ is a single-mode squeezing operatorm and $\phi_k$ parameterize the mode where each photon subtraction is performed. The states in Eq.~(\ref{modelSt}) can be rewritten as $|\psi_m\rangle=S_1(\xi_1)S_2(\xi_2)|\widetilde{\psi}_m\rangle$, where the core state $|\widetilde{\psi}_m\rangle$ results from the commutation between the linear combination of annihilation operators $a_1\cos \phi_k+a_2 \sin \phi_k$ and the squeezing operators $S_i(\xi_i)$. Further, we limit ourselves to the cases of single ($m=1$), or two ($m=2$) photon subtractions. 

\subsubsection{Single photon subtracted states}
\begin{figure*}[t]
	%\vspace{-1cm}
	\includegraphics[width=16cm]{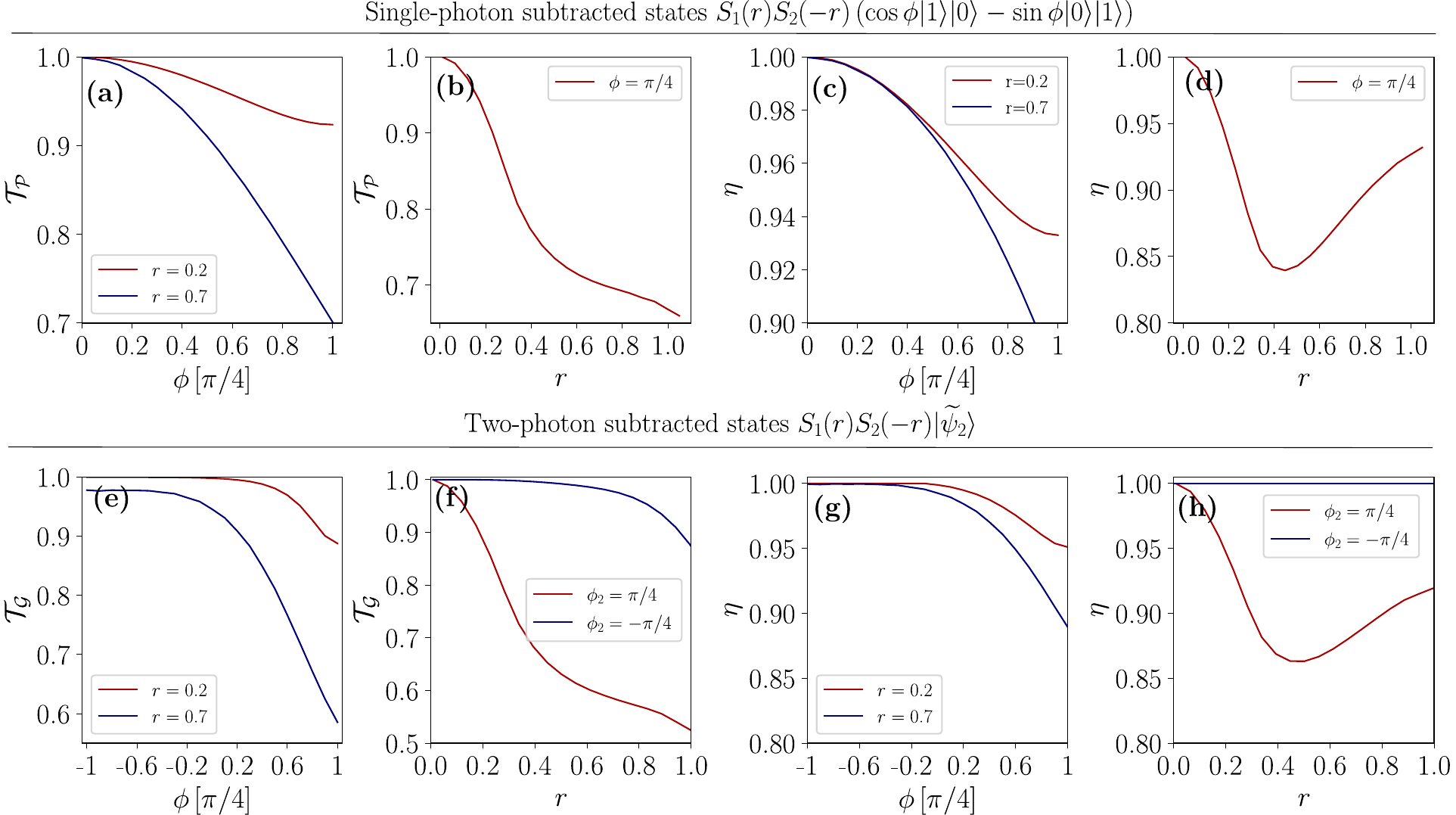}
	%\vspace{-1cm}
	\caption{ a) Threshold for the certification of mode-intrinsic entanglement using single-photon subtracted states $|\psi_1\rangle\propto S_1(r)S_2(-r)\left(\cos \phi |1\rangle|0\rangle-\sin \phi |0\rangle|1\rangle\right)$ as target states with fixed $r=0.2$ (red) and $r=0.7$ (blue) and different $\phi$. b) Thresholds using the same target state as in a) but with given $\phi=\pi/4$,  as a function of $r$. c) The minimal transmission $\eta$ of a lossy channel preserving the certification of mode-intrinsic entanglement of the state $|\psi_1\rangle$, with respect to the angle $\phi$ and given $r=0.2$ (red) and $r=0.7$ (blue). d) The minimal transmission $\eta$ of a lossy channel depicted for states $|\psi_1\rangle$ with respect to $r$ and fixed $\phi=\pi/4$. e) Threshold for genuine non-Gaussian entanglement based on the fidelity with two-photon subtracted state $|\psi_2\rangle=S_1(r)S_2(-r)|\widetilde{\psi}_2\rangle$ as a function of the angle $\phi_2$, defining the mode of the second photon subtraction, but with fixed angle $\phi_1=\pi/4$, determining the mode of first photon subtraction, and with given squeezing $r=0.2$ (red), $r=0.7$ (blue). f) Thresholds using the same target state as in e) but with fixed $\phi_2=\pi/4$ (red), and $\phi_2=-\pi/4$ (blue) a as a function of $r$. g) The minimal transmission $\eta$ of a lossy channel preserving the certification of genuine non-Gaussian entanglement of the state $|\psi_2\rangle$ are shown with respect to $\phi_2$ for fixed $r=0.2$ (red), $r=0.7$ (blue). h) Similar plot as in g) but with respect to $r$ and fixed $\phi_2=\pi/4$ (red), and $\phi_2=-\pi/4$ (blue). The minimal transmission $\eta$ for the case $\phi_2=-\pi/4$ is always greater than $0.99$.}
    \label{fig4}
\end{figure*}

The single-photon subtraction leads to the core state $|\widetilde{\psi}_1\rangle \propto \cos \phi \sinh \xi_1 |1\rangle|0\rangle+\sin \phi \sinh \xi_2 |0\rangle|1\rangle$, which is a passive separable state \cite{Lopetegui2025} and, consequently, the state $|\psi_1\rangle=S_1(\xi_1)S_2(\xi_2)|\widetilde{\psi}_1\rangle$ is always Gaussian separable \cite{Geometric_2025,chabaud_holomorphic_2022}. However, the squeezing operators acting on the state $|\widetilde{\psi}_1\rangle$ produce mode-intrinsic entanglement. To illustrate it, Fig.~\ref{fig4} a) depicts a threshold derived for the fidelity with the state $S_1(\xi_1)S_2(\xi_2)|\widetilde{\psi}_1\rangle$ with $\xi_2=-\xi_1=r$, as a function of the angle $\phi$ determining the mode where a photon is subtracted. We consider two different squeezing values $r=\{0.2,0.7\}$ and observe that for higher squeezing values the threshold drops significantly for any $\phi\neq 0$.  This drop seems to slow down for large squeezing, as we can observe in Fig.~\ref{fig4} b). The thresholds were computed using a Fock cutoff of 25, which is sufficient to correctly reproduce the state.\par 
In Fig.~\ref{fig4} c) and d), we show as well the loss tolerance for the certification of mode-intrinsic entanglement on those states, both as a function of $\phi$ (for two different values of $r$), and as a function of $r$ (for $\phi=\pi/4$). Interestingly, the loss tolerance is not monotonic as a function of $r$, but starts getting worse after a certain level of squeezing $r\sim 0.45$. \par
\subsubsection{Two photon subtracted states}
Here, we investigate two-photon subtracted states $ |\psi_2\rangle=S_1(\xi_1)S_2(\xi_2)|\widetilde{\psi}_2\rangle$ with $|\widetilde{\psi}_2\rangle=c_{0,2}|0\rangle|2\rangle+c_{2,0}|2\rangle|0\rangle+c_{1,1}|1\rangle|1\rangle+c_{0,0}|0\rangle|0\rangle$, where the amplitudes $c_{i,j}$ come from the commutation between the operators $a_i$ and the squeezing operators in Eq.~(\ref{modelSt}). For certain configurations of the parameters the core state $|\widetilde{\psi}_2\rangle$ is passive separable, yet in general it will be neither passive, nor Gaussian separable. 
%One such configuration is given by $\xi_1=\xi_2$, and $\phi_1=-\phi_2=\pi/4$. 
To demonstrate it, we analyse the certification of genuine non-Gaussian entanglement and mode-intrinsic entanglement for particular examples of two photon subtracted states.  \par
Fig.\ref{fig4} e), shows the behavior of the fidelity threshold against the choice of the mode where the second subtraction happens, parametrized by the angle $\phi_2$, for the case where $\xi_1=-\xi_2=r$, and $\phi_1=\pi/4$. We show the behavior for two different squeezing values, $r=0.2$, and $r=0.7$. We can observe that in both cases the threshold is high for the angle $\phi_2$ close to $-\pi/4$, which implies that it is difficult to use those states as targets for any certification. This is an interesting and somehow unexpected result, yet we should keep in mind that the fidelity cannot be seen as a measure of entanglement, but rather as a witness, and thus we cannot assume that these states have less genuine non-Gaussian entanglement than the ones obtained with $\phi_2=\pi/4$, \ie when the two photon subtractions happen in the same mode. The high values of fidelity obtained correspond to choosing a product of local states that have a similar parity as a squeezed states. This is because the effective operation that results from two subtractions in orthogonal modes, is given by $\hat a_1^2-\hat a_2^2$, which preserves the parity of the input squeezed states. 
In Fig.\ref{fig4} f), we show the behavior of the threshold against the squeezing value for $\phi_2=\pm \pi/4$. We observe that the thresholds in both cases  decrease with the squeezing $r$. For the choice of $\phi_2=-\pi/4$, the threshold remains consistently large, and thus difficult to beat in practice, for all values of squeezing considered.  
Finally, in Fig.\ref{fig4} g) we show the behavior of the loss tolerance threshold for certification based on the fidelity to the corresponding pure state, as a function of the choice of mode for the second subtraction, for $r=0.2$ and $r=0.7$. In Fig.\ref{fig4} h), a similar plot is shown, as a function of the squeezing value, for the choices $\phi_2=\pm\pi/4$. The behavior is similar to the one for single photon subtracted states. For the case $\phi_2=-\pi/4$, the tolerance to losses is quite low, which adds evidence on the difficulty of witnessing genuine non-Gaussian entanglement for this choice of parameters using this method. \par 
In all the cases that we have shown for two photon subtracted states, the threshold for the certification of genuine non-Gaussian entanglement and mode-intrinsic entanglement were the same. As a consequence, there is no clear separation between the loss threshold for the certification of mode-intrinsic and genuine non-Gaussian entanglement. Such separation would probably be observed in the examples we considered if, instead of fixing the target state for the certification, a more general optimization of the target state for each value of losses was performed. However, this is a hard task that is beyond the scope of the present work. 
%For some of the states, we show the variation of the fidelity to the pure state, as a function of losses, that act homogeneously on the two modes. Finally, in the rightmost plot in Fig.\ref{fig:two_photon_subtraction}, we analyse the effect of losses in the certification of genuine non-Gaussian entanglement on two photon subtracted states using their corresponding pure state as targets. In a similar trend as seen for other examples, losses not bigger than $6\%$ can be tolerated. 
\par

\section{Conclusion}
The certification of mode-intrinsic entanglement and genuine non-Gaussian entanglement identifies advanced forms of entanglement that can only be present in non-Gaussian states. We developed a procedure to derive certification criteria using fidelity with a target state. The approach relies on maximizing the fidelity over all Gaussian (or passive linear optics) evolutions of separable states. Importantly, the maximization over separable states can be performed efficiently, which makes the method computationally feasible even for arbitrarily complex target states. We demonstrated the computational power of the approach by deriving several criteria, using different states of interest as targets. We probe states of relevance for discrete variable, hybrid and purely continuous variable encoding.  For all cases we analysed the efect of losses on the potential certification of non-Gaussian entanglement. 
\par
Moreover, we considered the case of single and two photon subtracted states, which are within the reach of current quantum optical experiments and may provide an immediate testbed for our method. The results observed and tolerance to losses show that a certification based on fidelities might be within the reach of near term experiments, provided slight improvements on experimental losses. Nevertheless, a more detailed analysis should consider more general noise models. Furthermore, in our analysis of the loss tolerance we only consider a fixed target state: the state entering the loss channel. Nevertheless, it is possible that, as losses increase, the state gets closer in fidelity to a different mode-intrinsic entangled or genuine non-Gaussian entangled state. This implies that the loss tolerance could in principle be increased by an optimized choice of target state, which is in general a hard task for mixed states.

\begin{acknowledgments}
L.L. acknowledges the support from the Czech Science Foundation
(Grant No. 23-06015O). C.E.L.G., M.F., and M.W. acknowledge funding from the European Union’s Horizon Europe Framework Programme (EIC Pathfinder Challenge project Veriqub) under Grant Agreement No.\ 101114899. C.E.L.G. and M.W. are supported by Plan France 2030 through the project OQuLus (ANR-22-PETQ-0013). This project has received funding from the European Union’s Horizon Europe research and innovation programme under the Marie Skłodowska-Curie grant agreement No 101208849.
\end{acknowledgments}

%\begin{thebibliography}{99}
\bibliographystyle{quantum}
\bibliography{QNGEntanglementDoi.bib}

%\end{thebibliography}
\appendix

\section{Decomposition of two-mode Gaussian dynamics}
The Bloch-Messiah decomposition parametrizes any Gaussian dynamic $\widetilde{G}$ of two bosonic modes as:
\begin{equation}
	\widetilde{G}=U_{BS}(\tau_1)S_1(\xi_1)S_2(\xi_2)U_{BS}(\tau_2) D_1(\alpha_1)D_2(\alpha_2),
\end{equation}
where $U_{BS}(\tau)=\exp(\tau a_1 a_2^{\dagger}-\tau^* a_1^{\dagger}a_2)$ is the unitary operator corresponding to beam-splitter interaction, $D_i(\alpha)=\exp(\alpha a^{\dagger}_i-\alpha^* a_i)$ and $S_i(\xi)=\exp\left[\xi \left(a^{\dagger}_i\right)^2-\xi^* a_i^2\right]$ are the displacement operator and squeezing operators acting on the $i$th mode. We are interested in the entanglement that $\widetilde{G}$ produces when it acts on the state $|\widetilde{\psi}_1\rangle |\widetilde{\psi}_2\rangle$. Let us introduce the Gaussian dynamics
\begin{equation}
	G=U_{BS}(\tau_1)S_1(\xi_1)S_2(\xi_2)B_{BS}(\tau_2)
\end{equation}
without the displacement operators.
Since we have
\begin{equation}
	\widetilde{G}|\widetilde{\psi}_1\rangle |\widetilde{\psi}_2\rangle=G |\psi_1\rangle, |\psi_2\rangle,
\end{equation}
where $|\psi_i\rangle=D_i(\alpha_i)|\widetilde{\psi}_i\rangle$.
The operators $D_i(\alpha_i)$ do not increase the capability of the Gaussian evolution to produce entanglement, and therefore we ignore them further.

Here, we show that any Gaussian dynamics $G$ can be substituted by the operator
\begin{equation}
	G=U_{BS}(\tau_1)S_{1,2}(\xi)U_{BS}(\tau_2)S_1(\xi_1),
\end{equation}
where $S_{1,2}(\xi)=\exp(\xi a^{\dagger}b^{\dagger}-\xi^* a b)$ corresponds to two-mode squeezing operator. Thus, general Gaussian dynamics producing entanglement has the parameterization
\begin{equation}
	G=U_{BS}(\tau_1)S_{1,2}(\xi)U_{BS}(\tau_2).
\end{equation}

\textbf{Lemma 1}: The operators $S(\xi)$ and $U_{BS}(\tau)$  obey
\begin{equation}
	\begin{aligned}
		& U_{BS}(\widetilde{\tau})S_1(\xi_1)S_2(\xi_2)=R_1(\phi_1-\phi_2)S_1(\xi_1)S_2(\xi_2)\\
		& \times\exp\left(\tau a_1^{\dagger}a_2-\tau^* a_1 a_2^{\dagger}+\sigma a_1^{\dagger}a_2^{\dagger}-\sigma^* a_1 a_2\right)\\
        &\times R^{\dagger}_1(\phi_1-\phi_2),
	\end{aligned}
\end{equation}
where $R_1(\phi)=\exp\left(i\phi a^{\dagger}_1 a_1\right)$, $\tau=\widetilde{\tau}\cosh (|\xi_1|-|\xi_2|)$, $\sigma=\widetilde{\tau}\exp\left[i(\phi_1-\phi_2)\right]\sinh (|\xi_1|-|\xi_2|)$ and the phase $\phi_i$ is given by $\xi_i=|\xi|\exp(i\phi_i)$.

\textit{Proof}: We prove this lemma in Heisenberg picture that guarantees the identity
\begin{equation}\label{SM:HD}
	U f\left(a,a^{\dagger}\right) U^{\dagger}=f\left(U a U^{\dagger},U a^{\dagger} U^{\dagger}\right)
\end{equation}
for any function $f\left(a,a^{\dagger}\right)$ of the operators $a$ and $a^{\dagger}$ and an arbitrary unitary evolution $U$.
First, we employ the identity
\begin{equation}
	\begin{aligned}
		&U_{BS}(\tau)S_1(\xi_1)S_2(\xi_2)=\\
		&R_1(\phi) R_1^{\dagger}(\phi) U_{BS}(\tau)S_1(\xi_1)S_2(\xi_2)R_1(\phi) R_1^{\dagger}(\phi)=\\
		&R_1(\phi) U_{BS}(\widetilde{\tau})S_1(\widetilde{\xi}_1)S_2(\xi_2) R_1^{\dagger}(\phi),
	\end{aligned}
\end{equation}
where we obtain $\widetilde{\tau}=\tau \exp(i\phi)$ and $\widetilde{\xi}_1=\xi_1 \exp(2i\phi)$ due to Eq.~(\ref{SM:HD}).
We choose $\phi$ such that the complex parameters $\widetilde{\xi}_1$ and $\xi_2$
gain the same phase. 
Second, we consider the evolution
\begin{equation}
	S_i(\xi)a_i S_i^{\dagger}(\xi)=a_i\cosh 2|\xi|+a_i^{\dagger}e^{i\phi}\sinh 2|\xi|
\end{equation}
with $\xi=|\xi|e^{i\phi}$. Assuming $\xi_1=|\xi_1|\exp(i \phi)$ and $\xi_2=|\xi_2|\exp(i \phi)$, we obtain
\begin{equation}
	\begin{aligned}
		&U_{BS}(\tau)S_1(\xi_1)S_2(\xi_2)=\\
		&S_1(\xi_1)S_2(\xi_2) S_1^{\dagger}(\xi_1) S_2^{\dagger}(\xi_2) U_{BS}(\tau)S_1(\xi_1)S_2(\xi_2)=\\
		& S_1(\xi_1)S_2(\xi_2)
		\exp\left(\widetilde{\tau} a_1^{\dagger}a_2-\widetilde{\tau}^* a_1 a_2^{\dagger}+\sigma a_1^{\dagger}a_2^{\dagger}-\sigma^* a_1 a_2\right),
	\end{aligned}
\end{equation}
where $\sigma=\tau \exp(i\phi)\sinh(|\xi_1|-|\xi_2|)$ and $\widetilde{\tau}=\tau \cosh (|\xi_1|-|\xi_2|)$, which proves the lemma.

\textbf{Lemma 2}: Let us introduce parameters $\sigma$ and $\widetilde{\tau}$ that obey $|\sigma|<|\widetilde{\tau}|$. We obtain the identity
\begin{equation}
	\begin{aligned}
		& S_1(\xi)S_2(-\xi) \exp\left(\widetilde{\tau} a_1^{\dagger}a_2-\widetilde{\tau}^* a_1 a_2^{\dagger}+\sigma a_1^{\dagger}a_2^{\dagger}-\sigma^* a_1 a_2\right)=\\
		& U_{BS}(\tau) S_1(-\xi)S_2(\xi),
	\end{aligned}
\end{equation}
where $\sinh 2|\xi|=|\sigma|$ and $\tau=\widetilde{\tau}/\cosh 2|\xi|$.

\textit{Proof}: To prove this lemma, we follow analogously proof of Lemma 1.

\textbf{Theorem 1}: Any entanglement generating Gaussian dynamics is expressed as 
\begin{equation}
	G(\tau_1,\tau_2,\xi)=U_{BS}(\tau_1)S_{1,2}(\xi)U_{BS}(\tau_2),
\end{equation}
where $S_{1,2}(\xi)=\exp\left(\xi a_1^{\dagger}a_2^{\dagger}-\xi^* a_1 a_2\right)$ is a two-mode squeezing operator and $U_{BS}(\tau)=\exp\left(\tau a_1^{\dagger}a_2-\tau^* a_1 a_2^{\dagger}\right)$.

\textit{Proof}: Based on Bloch-Messiah decomposition \cite{Bloch1962,Braunstein2005}, we express a Gaussian evolution following as:
\begin{equation}\label{SM:BlochMessiah}
	G=U_{BS}(\tau_1)S_1(\xi_1)S_2(\xi_2)U_{BS}(\tau_2).
\end{equation}
We show that, for any given factorized state $|\psi_1\rangle |\psi_2\rangle$ and any given $G$, we obtain the state $|\widetilde{\psi}_1\rangle |\widetilde{\psi}_2\rangle$ that obeys the identity
\begin{equation}
	\begin{aligned}
			&G|\psi_1\rangle|\psi_2\rangle=\\
			&U_{BS}(\tau_1)S_{1,2}(\xi)U_{BS}(\tau_2)|\widetilde{\psi}_1\rangle|\widetilde{\psi}_2\rangle
	\end{aligned}
\end{equation}
for some $\tau_1$, $\tau_2$ and $\xi$.
We employ Lemma 1 and Lemma 2 to express the Gaussian operator in Eq.~(\ref{SM:BlochMessiah}) following as:
\begin{equation}
	\begin{aligned}
	& G(\tau_1,\tau_2,\xi_1,\xi_2)=R_1(\phi_2-\phi_1)U_{BS}(\tau_1-\pi/4)\\
	& \times U_{BS}(\tau=\pi/4)S_1(-\xi_2)S_2(\xi_2)U_{BS}(\tau=\pi/4)\\
	&\times U_{BS}(\tau_2-\pi/4)S_1(\widetilde{\xi}_1+\xi_2)R^{\dagger}_1(\phi_2-\phi_1),
	\end{aligned}
\end{equation}
where we assume $\xi_j=|\xi_j|\exp(i\phi_j)$ and $\widetilde{\xi}_1=|\xi_1|\exp(i\phi_2)$. Consequently, we gain
\begin{equation}\label{SM:theoremS2}
	\begin{aligned}
			& G |\psi_1\rangle|\psi_2\rangle =R_1(\phi_2-\phi_1)U_{BS}(\tau-\pi/4)\\
			&  \times S_{1,2}(\xi_2)U_{BS}(\tau-\pi/4)|\widetilde{\psi}_1\rangle|\psi_2\rangle
	\end{aligned}
\end{equation}
with $|\widetilde{\psi}_1\rangle=S_1(\widetilde{\xi}_1+\xi_2)R^{\dagger}_1(\phi_2-\phi_1)|\psi_1\rangle|\psi_2\rangle$ and $S_{1,2}(\xi_2)=U_{BS}(\tau=\pi/4)S_1(-\xi_2)S_2(\xi_2)U_{BS}(\tau=\pi/4)$. Further, we commute the operator $R_1(\phi_1-\phi_2)$ in Eq.~(\ref{SM:theoremS2}) to the right side, which leads to
\begin{equation}\label{SM:theoremS3}
	\begin{aligned}
		& G |\psi_1\rangle|\psi_2\rangle =U_{BS}(\widetilde{\tau})\\
		&  \times S_{1,2}(\widetilde{\xi}_2)U_{BS}(\widetilde{\tau})R^{\dagger}_1(\phi_2-\phi_1)|\widetilde{\psi}_1\rangle|\psi_2\rangle,
	\end{aligned}
\end{equation}
where $\widetilde{\tau}=(\tau-\pi/4)\exp\left[i(\phi_2-\phi_1)\right]$ and $\widetilde{\xi}=\xi \exp\left[i(\phi_2-\phi_1)\right]$. Finally, Eq.~(\ref{SM:theoremS3}) proves the theorem.

\textbf{Theorem 2}: Let $R_i(\phi)=\exp\left(i\phi a_i^{\dagger}a_i\right)$ denote the unitary operator of the phase shift in the $i$th mode. Any entanglement generating Gaussian dynamics is given by
\begin{equation}
	\begin{aligned}
		& G(\tau_1,\tau_2,\xi,\phi)\\
		& =R_1(\phi_1)R_2(\phi_2)U_{BS}(\tau_1)R_1(\phi)S_{1,2}(\xi)U_{BS}(\tau_2),
	\end{aligned}
\end{equation}
where $\tau_1$, $\tau_2$ and $\xi$ are real.

\textit{Proof:} Employing Theorem 1, we express a general entanglement generating Gaussian dynamics following as:
\begin{equation}
	G(\tau_1,\tau_2,\xi)=U_{BS}(\tau_1)S_{1,2}(\xi)U_{BS}(\tau_2)
\end{equation}
with complex $\tau_1$, $\tau_2$ and $\xi$. Since $R_i(\phi)a_i R_i^{\dagger}(\phi)=a_i \exp(i\phi)$, we obtain the identities
\begin{equation}\label{SM:rot}
	\begin{aligned}
		& R_1(\phi_1) R_2(\phi_2) U_{BS}(\tau)R_1^{\dagger}(\phi)R_2^{\dagger}(\phi_2)=U_{BS}\left(\tau e^{i\phi_1-i\phi_2}\right)\\
		& R_1(\phi_1) R_2(\phi_2) S_{1,2}(\xi)R_1^{\dagger}(\phi)R_2^{\dagger}(\phi_2)=S_{1,2}\left(\xi e^{i \phi_1+i \phi_2}\right).
	\end{aligned}
\end{equation}
Eq.~(\ref{SM:rot}) implies that we can exploit the operator $R_i(\phi)$ to change the phase of the parameters $\tau_1$, $\tau_2$ and $\xi$ such that we arrive at
\begin{equation}
	\begin{aligned}
		& G(\tau_1,\tau_2,\xi)\\
		& =R_1(\phi_1)R_2(\phi_2)U_{BS}(|\tau_1|)R_1(\phi)S_{1,2}(|\xi|)U_{BS}(|\tau_2|)\\
		&\times R_1(\widetilde{\phi}_1)R_2(\widetilde{\phi}_2),
	\end{aligned}
\end{equation}
for some properly chosen $\phi$, $\phi_1$, $\phi_2$, $\widetilde{\phi}_1$ and $\widetilde{\phi}_2$.

\section{Analytical formulas for parametrization of entanglement-generating Gaussian dynamics}
Here, we calculate the overlap $o_{k,l,m,n}\equiv \langle n|\langle m|U(\tau_1)R_1(\phi) S_{1,2}(\xi)U(\tau_2)|k\rangle|l\rangle$. We express $o_{k,l,m,n}$ following as:
\begin{equation}\label{SM:overlap1}
	\begin{aligned}
			& o_{k,l,m,n}=\frac{1}{\sqrt{k!l!m!n!}}\langle 0|\langle 0|a_1^m a_2^n U(\tau_1)S_{1,2}(\xi)U(\tau_2)\\
			& \times \left(a_1^{\dagger}\right)^k\left(a_2^{\dagger}\right)^l|0\rangle|0\rangle=\\
			&\lim_{\lambda_k,\lambda_l,\lambda_m,\lambda_n \rightarrow 0}\frac{1}{\sqrt{k!l!m!n!}}\partial_{\lambda_k}^k\partial_{\lambda_l}^l\partial_{\lambda_m}^m\partial_{\lambda_n}^n \langle 0|\langle 0|e^{\lambda_k a_1+\lambda_l a_2}\\
			&U(\tau_1)S_{1,2}(\xi)U(\tau_2)e^{\lambda_m a_1^{\dagger}+\lambda_2 a_2^{\dagger}}|0\rangle|0\rangle
	\end{aligned}
\end{equation}
and use the commutation rules for gaining
\begin{equation}\label{SM:overlap2}
	\begin{aligned}
		&\langle 0|\langle 0|e^{\lambda_k a_1+\lambda_l a_2}U(\tau_1)S_{1,2}(\xi)U(\tau_2)e^{\lambda_m a_1^{\dagger}+\lambda_2 a_2^{\dagger}}|0\rangle|0\rangle=\\
		&\langle 0|\langle 0|e^{f_1 a^{\dagger}_1+f_2 a^{\dagger}_2+g_1 a_1+g_2 a_2}S_{1,2}(\xi)|0\rangle|0\rangle,
	\end{aligned}
\end{equation}
where the coefficients $f_i$ and $g_i$ with $i=1,2$ depend on $\tau_1$, $\tau_2$, $\xi$, $\lambda_k$, $\lambda_l$, $\lambda_m$ and $\lambda_n$. We use the identity
\begin{equation}\label{SM:overlap3}
	\begin{aligned}
		&\langle 0|\langle 0|e^{f_1 a^{\dagger}_1+f_2 a^{\dagger}_2+g_1 a_1+g_2 a_2}S_{1,2}(\xi)|0\rangle|0\rangle=\\
		& \langle 0|\langle 0|e^{f_1 a^{\dagger}_1+f_2 a^{\dagger}_2}e^{-\left(f_1 g_1+f_2 g_2\right)/2+g_1 a_1+g_2 a_2}S_{1,2}(\xi)|0\rangle|0\rangle\\
		&=e^{-\left(f_1 g_1+f_2 g_2\right)/2}\langle 0|\langle 0|e^{g_1 a_1}e^{g_2 a_2}S_{1,2}(\xi)|0\rangle|0\rangle.
	\end{aligned}
\end{equation}
Further, we consider
\begin{equation}
	\begin{aligned}
		&\langle 0|\langle 0|e^{g_1 a_1}e^{g_2 a_2}S_{1,2}(\xi)|0\rangle|0\rangle=\\
		&\frac{1}{\cosh |\xi|}\langle 0|\langle 0|\sum_{k=0}^{\infty}\frac{\left(g_1 a_1\right)^k}{k!}\sum_{l=0}^{\infty}\frac{\left(g_2 a_2\right)^l}{l!}\\
		&\times \sum_{n=0}^{\infty}\frac{\left(\tanh \xi a_1^{\dagger} a_2^{\dagger}\right)^n}{n!}|0\rangle|0\rangle=\\
		&\frac{1}{\cosh\xi}\sum_{n=0}^{\infty}\frac{\left(g_1 g_2 \tanh\xi\right)^n}{n!}=\frac{1}{\cosh \xi}e^{g_1 g_2 \tanh \xi}.
	\end{aligned}
\end{equation}
Combining Eq.~(\ref{SM:overlap1}), (\ref{SM:overlap2}) and (\ref{SM:overlap3}), we obtain
\begin{equation}\label{SM:analFinal}
	\begin{aligned}
		&o_{k,l,m,n}=\lim_{\lambda_k,\lambda_l,\lambda_m,\lambda_n \rightarrow 0}\frac{1}{\sqrt{k!l!m!n!}}\times \\
		&\partial_{\lambda_k}^k\partial_{\lambda_l}^l\partial_{\lambda_m}^m\partial_{\lambda_n}^n\frac{1}{\cosh \xi}e^{-\left(f_1 g_1+f_2 g_2\right)/2+g_1 g_2 \tanh \xi}.
	\end{aligned}
\end{equation}
The differentiation in Eq.~(\ref{SM:analFinal}) allows us to derive the corresponding analytical formula for the overlap $o_{k,l,m,n}$.

\end{document}